\documentclass[aps,prl,twocolumn,superscriptaddress,showpacs]{revtex4}
\usepackage{amssymb}
\usepackage{txfonts}
\usepackage{tipa}

\usepackage{graphicx}
\usepackage{dcolumn}
\usepackage{bm}

\begin{document}


\title{Neutron diffraction study on phase transition and thermal expansion of SrFeAsF}

\author{Y. Xiao}
\email[y.xiao@fz-juelich.de]{}
\affiliation{Institut fuer Festkoerperforschung, Forschungszentrum
Juelich, D-52425 Juelich, Germany}

\author{Y. Su}
\affiliation{Juelich Centre for Neutron Science, IFF,
Forschungszentrum Juelich, Outstation at FRM II, Lichtenbergstrasse
1, D-85747 Garching, Germany}

\author{R. Mittal}
\affiliation{Juelich Centre for Neutron Science, IFF,
Forschungszentrum Juelich, Outstation at FRM II, Lichtenbergstrasse
1, D-85747 Garching, Germany}
\affiliation{Solid State Physics
Division, Bhabha Atomic Research Centre, Trombay, Mumbai 400 085,
India}

\author{T. Chatterji}
\affiliation{Juelich Centre for Neutron Science, IFF,
Forschungszentrum Juelich, Outstation at Institut Laue-Langevin, BP
156, 38042 Grenoble Cedex 9, France}

\author{T. Hansen}
\affiliation{Institut Laue-Langevin, BP 156, 38042 Grenoble Cedex 9,
France}

\author{S. Price}
\affiliation{Juelich Centre for Neutron Science, IFF,
Forschungszentrum Juelich, Outstation at FRM II, Lichtenbergstrasse
1, D-85747 Garching, Germany}

\author{C. M. N. Kumar}
\affiliation{Institut fuer Festkoerperforschung, Forschungszentrum
Juelich, D-52425 Juelich, Germany}

\author{J. Persson}
\affiliation{Institut fuer Festkoerperforschung, Forschungszentrum
Juelich, D-52425 Juelich, Germany}

\author{S. Matsuishi}

\affiliation{Frontier Research Center, Tokyo Institute of
Technology, 4259 Nagatsuta-cho, Midori-ku, Yokohama 226-8503, Japan}

\author{Y. Inoue}

\affiliation{Frontier Research Center, Tokyo Institute of
Technology, 4259 Nagatsuta-cho, Midori-ku, Yokohama 226-8503, Japan}

\author{H. Hosono}

\affiliation{Frontier Research Center, Tokyo Institute of
Technology, 4259 Nagatsuta-cho, Midori-ku, Yokohama 226-8503, Japan}

\author{Th. Brueckel}
\affiliation{Institut fuer Festkoerperforschung, Forschungszentrum
Juelich, D-52425 Juelich, Germany} \affiliation{Juelich Centre for
Neutron Science, IFF, Forschungszentrum Juelich, Outstation at FRM
II, Lichtenbergstrasse 1, D-85747 Garching, Germany}
\affiliation{Juelich Centre for Neutron Science, IFF,
Forschungszentrum Juelich, Outstation at Institut Laue-Langevin, BP
156, 38042 Grenoble Cedex 9, France}

\date{\today}

\begin{abstract}

The magnetic ordering and crystal structure of iron pnictide SrFeAsF
was investigated by using neutron powder diffraction method. With
decreasing temperature, the tetragonal to orthorhombic phase
transition is found at 180 K, while the paramagnetic to
antiferromagnetic phase transition set in at 133 K. Similar to the
parent compound of other iron pnictide system, the striped Fe
magnetism is confirmed in antiferromagnetic phase and the Fe moment
of 0.58(6) $\mu$$_B$ aligned along long \emph{a} axis. The thermal
expansion of orthorhombic phase of SrFeAsF is also investigated.
Based on the Gr\"{u}neisen approximation and Debye approximation for
internal energy, the volume of SrFeAsF can be well fitted with Debye
temperature of 347(5) K. The experimental atomic displacement
parameters for different crystallographic sites in SrFeAsF are
analyzed with Debye model. The results suggested that the expansion
of FeAs layers plays an important role in determining the thermal
expansion coefficient.

\end{abstract}

\maketitle

High temperature superconductivity has recently been discovered in
carriers doped iron pnictide compounds
\cite{Kamihara,Chen1,Takahashi,Rotter}. Till now, the highest
\emph{T}$_c$ attained is 57.4 K in the electron doped '1111'
compound Ca$_{0.4}$Na$_{0.6}$FeAsF \cite{Cheng}, while for the so
called '122' compound the highest \emph{T}$_c$ of 38 K is reached in
the hole doped Ba$_{0.6}$K$_{0.4}$Fe$_2$As$_2$ \cite{Rotter}. It is
generally believed that the superconductivity in iron pnictides is
unlikely due to simple electron-phonon coupling, as demonstrated
from extensive studies of phonon dynamics \cite{Mittal1,Mittal2}.
Magnetism seems to play a crucial role in the appearance of
superconductivity and AFM spin fluctuations have thus been suggested
to be a possible pairing mechanism \cite{Christianson}.

The spin density wave (SDW) transition, which is associated with
long range antiferromagnetic (AFM) order of Fe moments, is shown in
undoped FeAs-based compounds \cite{Cruz,Zhao1,Huang,Su1}. Band
structural calculations suggested that the SDW is driven either by
Fermi surface nesting between electron and hole pockets or by the
magnetic exchange coupling between local moments
\cite{Mazin1,Yildirim,Si,Fang}. Furthermore, the SDW transition is
found to be accompanied by the tetragonal to orthorhombic (T-O)
structural phase transition in the 122 family and preceded by the
T-O transition for the 1111 family \cite{Cruz,Zhao1,Huang,Su1}. It
is generally accepted that the crystal and magnetic structure in
undoped FeAs compound are intimately coupled. Theoretical studies
suggested that the structural phase transition is driven by the AFM
stripe magnetism directly \cite{Yildirim,Fang,Mazin2}. Therefore, it
is important to investigate the relationship between structural
properties and AFM ordering of the undoped FeAs compound, especially
for the 1111 compound, in which the structural phase transition
always occurred at higher temperature than SDW transition
temperature.

As a new parent phase of iron pnictide family, the SrFeAsF also
exhibited prominent anomaly in magnetic and electric property
measurements \cite{Matsuishi1,Tegel,Han}. However, detailed studies
have not been carried out to investigate the magnetic and structural
transformation in SrFeAsF. In this paper, we investigated the
structural and magnetic phase transition as well as thermal
expansion in SrFeAsF compound by using the neutron powder
diffraction (NPD) method. Both structural and magnetic phase
transitions are clarified; the temperature dependence of atomic
displacement parameters and thermal expansion are well modeled based
on the experimental results. The SrFeAsF polycrystalline sample was
synthesized by a solid state reaction method as described in Ref
\cite{Matsuishi1} with impurity phases (CaF$_2$ and Fe$_2$O$_3$) of
less than 1\%. The neutron powder diffraction measurements were
performed on the high flux powder diffractometer D20 at Institut
Laue Langevin (Grenoble, France). A Ge (115) monochromator was used
to produce a monochromatic neutron beam of wavelength 1.88 $\, $\AA.
The sample was loaded in a vanadium sample holder and then installed
in the liquid helium cryostat that can generate temperature down to
2 K. The program FULLPROF \cite{Carvajal} was used for the Rietveld
refinement of the crystal and the magnetic structures of the
compounds.

\begin{figure}
\includegraphics[width=8.5cm,height=9cm]{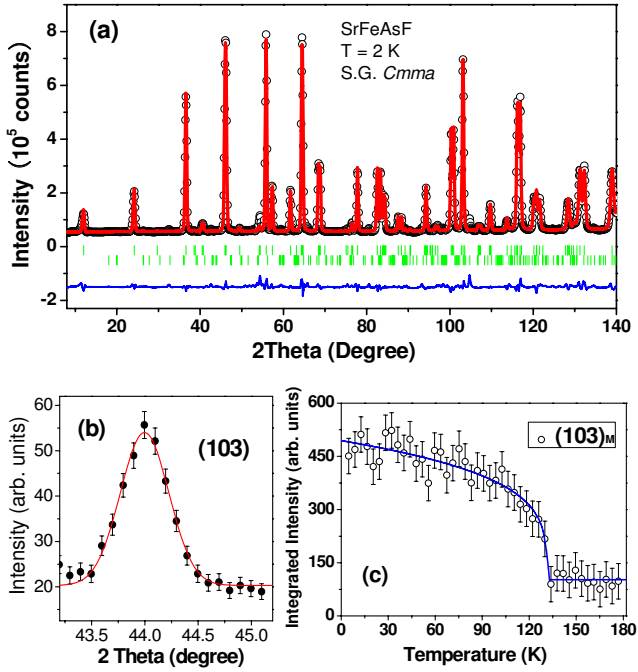}
\caption{\label{fig:epsart} (Color online) (a)Neutron powder
diffraction pattern of SrFeAsF at 2 K. The cirlces represent the
observed intensities, the solid line is the calculated pattern. The
difference between the observed and calculated intensities is shown
at the bottom. The vertical bars indicate the expected nuclear and
magnetic Bragg reflection positions. (b) The (103)$_M$ magnetic
reflections obtained by subtracting the NPD pattern measured at 240
K from the pattern measured at 2 K. (c) Temperature dependence of
the integrated intensity of (103)$_M$ magnetic Bragg reflection. The
solid curve is fit to the power law.}
\end{figure}

Similar to the CaFeAsF compound, the SrFeAsF also crystallized in
orthorhombic structure with space group \emph{Cmma} at 2 K.
Refinement of the neutron diffraction pattern gave lattice
parameters of \emph{a} = 5.6689(1) $\, $\AA$^{}$, \emph{b} =
5.6260(1) $\, $\AA$^{}$ and \emph{c} = 8.9325(2) $\, $\AA$^{}$ at 2
K. Both lattice parameters and unit-cell volume \emph{V} are larger
than that in CaFeAsF, which is consistent with the fact that the
atomic radius of Sr is larger than that of Ca. The refined neutron
powder diffraction pattern of SrFeAsF at 2 K is shown in Fig. 1(a).
Apart from the nuclear Bragg reflections, the magnetic peaks can be
fitted by using the same AFM structural model as in CaFeAsF
\cite{Xiao1} for Fe magnetism and the Fe moment is deduced to be
0.58(6) $\mu$$_B$. It seems that the AFM configuration of Fe spins
and the small magnitude of moment are a common feature for all FeAs
systems. The origin of that small iron moment in these compounds was
explained theoretically as the result of the itinerant character of
iron spins \cite{Mazin1} or the nearest and next nearest neighbor
superexchange interactions between Fe ions which give rise to a
frustrated magnetic ground state \cite{Yildirim}. The temperature
dependence of integrated intensity of (103)$_M$ magnetic reflection
was plotted in Fig. 1(c) to obtain the AFM phase transition
temperature. By fitting the ordering parameter with power law, the
Fe spins are found to order below \emph{T}$_N$ = 133(3) K. The onset
of the structural phase transition from \emph{P4/nmm} to \emph{Cmma}
takes place at \emph{T}$_S$ = 180(2) K as indicated by the evolution
of (220)$_O$ tetragonal reflection with temperature [Fig. 2(a)]. The
\emph{T}$_S$ is in accordance with the structural phase transition
observed by other methods \cite{Matsuishi1,Tegel,Han}.

The variation of lattice parameters can be obtained through the
refinement of NPD patterns. As shown in Fig. (b) and (c), the
lattice parameters \emph{a} and \emph{b} deviated each other when
the temperature is lower than the structural phase transition
temperature, while the lattice parameter \emph{c} decreased
gradually with decrease of temperature. The structural distortion
will result in an inequivalent of nearest- and next-nearest-neighbor
coupling (\emph{J}$_1$ and \emph{J}$_2$) in FeAs layers and lift the
magnetic frustration. The collinear AFM spin configuration is
stabilized as the ground state when \emph{J}$_2$ $>$ \emph{J}$_1$/2
\cite{Yildirim,Fang}. Within the Heisenberg magnetic exchange model,
it is established that the difference between the structural and AFM
transition temperature is strongly correlated with the ratio between
out-of-plane magnetic exchange coupling \emph{J}$_z$ and
\emph{J}$_2$ \cite{Fang,Luo}. To our knowledge, the difference
between \emph{T}$_S$ and \emph{T}$_N$ in SrFeAsF ($\sim$ 47 K) is
the largest among all known iron pnictides. Therefore, the big
temperature difference between \emph{T}$_s$ and \emph{T}$_N$
indicates the smallest \emph{J}$_z$/\emph{J}$_2$ ratio in SrFeAsF.

\begin{figure}
\includegraphics[width=8.5cm,height=6.5cm]{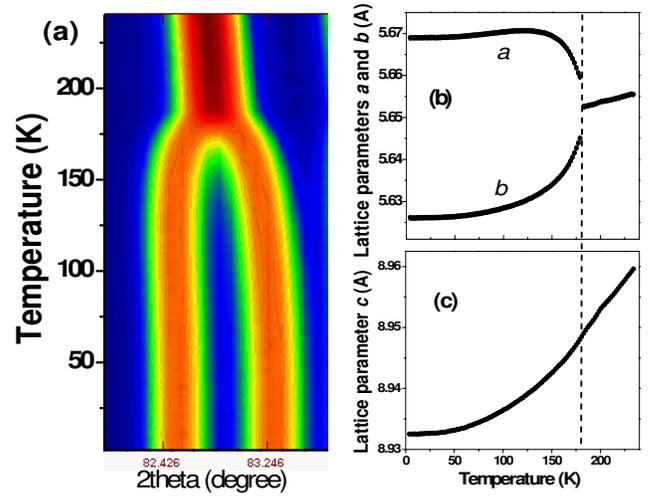}
\caption{\label{fig:epsart} (Color online) (a) Evolution of
(220)$_T$ reflections with the change of temperature. The structural
phase transition is clearly revealed by the splitting of tetragonal
(220)$_T$ into orthorhombic (400)$_O$ and (040)$_O$ reflections.
(b)(c) Temperature dependence of the estimated lattice parameter
\emph{a}, \emph{b} and \emph{c} for both tetragonal and orthorhombic
phases.}
\end{figure}

The detailed structural information for SrFeAsF at 2 K and 240 K, as
obtained from NPD data, are given in Table I. The variation of
volume with temperature over measured temperature range is shown in
Fig. 3. To model the experimental data of the unit cell volume, we
followed the approach of the Gr\"{u}neisen approximation for the
zero pressure equation of state, in which the effects of thermal
expansion are considered to be equivalent to elastic strain
\cite{Wallace}. Thus, the temperature dependence of the volume can
be described by \emph{V}(\emph{T}) =
$\gamma$\emph{U}(\emph{T})/\emph{K}$_0$ + \emph{V}$_0$, where
$\gamma$ is a Gr\"{u}neisen parameter, \emph{K}$_0$ is the
incompressibility and \emph{V}$_0$ is the volume at \emph{T} = 0 K.
By adopting the Debye approximation, the internal energy
\emph{U}(\emph{T}) is given by:
\begin{equation}
{\emph{U}(T)}= 9Nk_BT
 \left(
   \frac{T}{\theta_D}
 \right)^3\int_0^{\theta_D/T}\frac{x^3}{e^x-1}\mathrm{d}x
\end{equation}
where \emph{N} is the number of atoms in the unit cell, \emph{k}$_B$
is the Boltzmann's constant and $\theta$$_D$ is the Debye
temperature. Above model exhibits a good fit to volume variation as
indicated by the solid line in Fig. 3. We obtained following
physical parameters: $\theta$$_D$ = 347(5) K, \emph{V}$_0$ =
284.89(2) $\, $\AA$^{3}$ and $\gamma$/\emph{K}$_0$ =
1.57(3)$\times$10$^{-9}$ Pa$^{-1}$. The value for the Debye
temperature as determined from neutron experiments is in reasonable
agreement with those determined from specific heat measurements,
$\theta$$_D$ = 339(1) in Ref\cite{Tegel}. Note that the Debye
temperature approximates only the acoustical phonons at low
temperature. The thermal expansion coefficient $\alpha$(\emph{T})
which corresponds to the first derivative of \emph{U}(\emph{T})
divided by the unit cell volume at that temperature is plotted in
the inset of Fig. 3(a) as the solid dots. The solid curve correspond
to the calculated value of $\alpha$(\emph{T}) from
\emph{V}(\emph{T}).

\begin{table}
\caption{\label{tab:table1} Refined results of the crystal and
magnetic structures for SrFeAsF at 2 and 240 K. The atomic positions
for space group \emph{Cmma}: Sr(4\emph{g})(0,0.25,z),
Fe(4\emph{b})(0.25,0,0.5), As(4\emph{g})(0,0.25,z),
F(4\emph{a})(0.25,0,0); for \emph{P4/nmm}:
Sr(2\emph{c})(0.25,0.25,z), Fe(2\emph{b})(0.75,0.25,0.5),
As(2\emph{c})(0.25,0.25,z), F(2\emph{a})(0.75,0.25,0).}
\begin{ruledtabular}
\begin{tabular}{lll}
Temperature & 2 K & 240 K\\
\hline

Space group & \emph{Cmma} & \emph{P4/nmm}\\

\emph{a} \, ($\, $\AA$^{}$)&5.6689(1)&3.9996(1)\\
\emph{b} \, ($\, $\AA$^{}$)&5.6260(1)&3.9996(1)\\
\emph{c} \, ($\, $\AA$^{}$)&8.9325(2)&8.9618(4)\\
\emph{V} \, ($\, $\AA$^{3}$)&284.89(2)&143.36(2)\\

Sr \\
\quad \quad \, $\emph{z}$&0.1584(1)&0.1583(2)\\
\quad \quad \emph{B}($\,$\AA$^{2}$)&0.05(4)&0.44(4)\\

Fe \\
\quad \quad \emph{B}($\,$\AA$^{2}$)&0.04(2)&0.29(2)\\
\quad \quad \emph{M}$_a$($\mu$$_B$)&0.58(6)&\\

As \\
\quad \quad \, $\emph{z}$&0.6525(1)&0.6515(1)\\
\quad \quad \emph{B}($\,$\AA$^{2}$)&0.05(4)&0.33(4)\\

F \\
\quad \quad \emph{B}($\,$\AA$^{2}$)&0.13(3)&0.38(3)\\

Bondlength($\,$\AA)\\

Fe-Fe& 2.834(2)$\times$2& 2.828(2)$\times$4\\
&2.813(1)$\times$2&\\

Fe-As& 2.417(2)$\times$4& 2.417(2)$\times$4\\

Bondangle($\textordmasculine$)\\

Fe-As-Fe & 71.168(3)$\times$2&71.608(3)$\times$4\\
&71.794(3)$\times$2&111.653(3)$\times$2\\
&111.394(3)$\times$2&\\

\emph{R$_p$}&3.43&4.98\\
\emph{R$_{wp}$}&5.09&7.58\\
$\chi$$^2$&3.53&5.63\\

\end{tabular}
\end{ruledtabular}
\end{table}

\begin{figure}
\includegraphics[width=8.5cm,height=14cm]{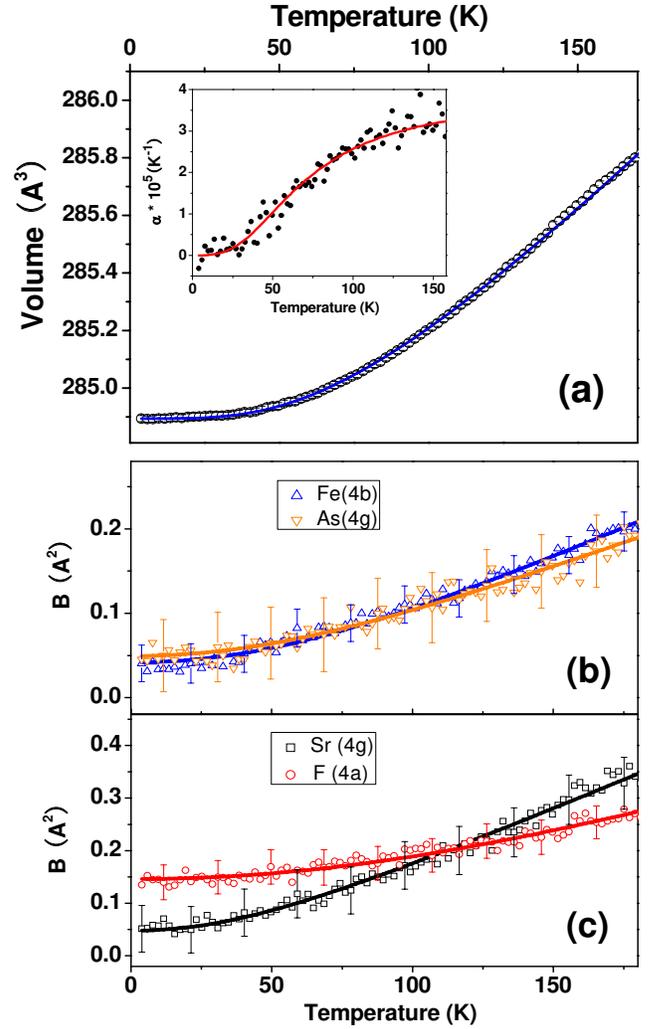}
\caption{\label{fig:epsart} (Color online) (a) Temperature
dependence of the volume of unit cell of SrFeAsF. The solid line is
fit of the Debye model as described in the text. The inset shows the
thermal expansion constant of volume as determined from the fit of
the Debye model in comparison to the experimental data. (b)(c)
Isotropic thermal parameters for different atomic sites in SrFeAsF.}
\end{figure}

As a basic crystallographic parameter, the atomic displacement
parameter (ADP) can reflect the atomic thermal motion and provide
useful information on the thermal properties of materials
\cite{Gross,Sales}. Therefore, the ADPs of different
crystallographic sites in SrFeAsF was also investigated. For the low
temperature \emph{Cmma} phase, the space group symmetries of all
special positions [Sr(4\emph{g}), Fe(4\emph{b}), As(4\emph{g}) and
F(4\emph{a})] allow three independent ADP elements. Based on
approximate classification, the isotropic ADP
$\langle$\emph{U}$_{iso}^2$$\rangle$ was applied by assuming that
the displacements are the same in all directions. The thermal
parameter \emph{B}$_{iso}$ which is obtained assuming isotropic
thermal motions of atoms are evaluated using the relation
\emph{B}$_{iso}$ = 8$\pi$$^2$$\langle$\emph{U}$_{iso}^2$$\rangle$ +
\emph{B}$_{sta}$, where \emph{B}$_{sta}$ is the static component of
the thermal parameter caused by the presence of a certain amount of
static disorder in compounds. Fig. 3(b) and Fig. 3(c) show the
temperature variation of experimental thermal parameters for all
four atomic sites of SrFeAsF. Debye model was adopted to describe
the ADP as a function of temperature:
\begin{equation}
{\langle U_{iso}^2\rangle}= \frac{3\hbar^2T}{mk_B\theta_D^2}
 \left(\Phi(\theta_D/T)+
   \frac{\theta_D}{4T}
 \right)
\end{equation}
where $\Phi(\theta_D/T)$ is given by
\begin{equation}
{\Phi(\theta_D/T)}=
   \frac{T}{\theta_D}
\int_0^{\theta_D/T}\frac{x}{e^x-1}\mathrm{d}x
\end{equation}

The fitted results for the thermal parameters based on above
equations are plotted in Fig. 3(b) and Fig. 3(c) as the solid lines
in comparison with the experimental results. The Debye temperatures
evaluated for different atomic species are Sr [238(6)K], Fe
[365(6)K], As [348(6)K] and F [611(6)K]. The averaged mass weighted
value for the Debye temperature is obtained to be 332(10) K. The
corresponding Debye frequencies for Sr, Fe, As, F sites are 4.9(1),
7.6(1), 7.2(1) and 12.7(1) THz respectively. The small difference
between the Debye frequencies for Fe and As revealed the similar
range of phonon frequencies for those two sites. The range of phonon
frequencies obtained from our analysis is in very good agreement
with our recent inelastic neutron scattering and \emph{ab initio}
phonon calculation results \cite{Mittal3} for SrFeAsF indicating the
reliability of our analysis.

\emph{ab initio} electronic structure calculations for SrFeAsF
\cite{Nebrasov} shows that it possessed essentially the same band
dispersions in the vicinity of the Fermi surface as in other 1111
compounds, such as LaFeAsO and CaFeAsF. Furthermore, the electronic
states near the Fermi surface are dominated by contributions from Fe
and As, which indicates that the FeAs layers are playing an
important role in magnetism and superconductivity of the compound.
From our thermal parameter analysis, we also notice that the mean
value of the Debye temperature for Fe and As [355(8) K] in SrFeAsF
agrees well with the Debye temperature deduced from the analysis of
thermal expansion curve [Fig. 3(a), $\theta$$_D$ = 347(5) K ].
Therefore, the thermal expansion of FeAs layers is supposed to be
the main contribution to the thermal expansion of volume.

In summary, the structural and magnetic phase transition as well as
the thermal expansion of SrFeAsF was investigated by using the
neutron diffraction method. The onset of tetragonal to orthorhombic
phase transition is at 180 K while the paramagnetic to
antiferromagnetic phase transition takes place at 133 K. SrFeAsF
exhibites the largest difference between \emph{T}$_S$ and
\emph{T}$_N$ among all known iron pnictides. The striped AFM
arrangement of Fe spins is confirmed by refinement of neutron data.
The magnitude of Fe moment aligned along \emph{a} direction is
deduced to be 0.58(6) $\mu$$_B$. The lattice volume of low
temperature orthorhombic phase can be modeled well in frame of
Gr\"{u}neisen approximation with a Debye approximation for the
internal energy. The analysis on atomic displacement parameters of
different crystallographic sites suggested that the vibration of Fe
and As is coupled in FeAs layers and they contribute mainly to the
thermal expansion of volume in SrFeAsF.

\appendix

\end{document}